\documentclass{emulateapj}
\usepackage{graphicx}
\usepackage{epstopdf}
\usepackage{epsfig}

\newcommand{\avg}[1]{\ensuremath{\langle #1 \rangle}}
\newcommand{\bma}{\begin{math}}
\newcommand{\ema}{\end{math}}
\newcommand{\beq}{\begin{equation}}
\newcommand{\eeq}{\end{equation}}
\newcommand{\beqa}{\begin{eqnarray}}
\newcommand{\eeqa}{\end{eqnarray}}
\newcommand{\bc}{\begin{center}}
\newcommand{\ec}{\end{center}} 
\newcommand{\bit}{\begin{itemize}}
\newcommand{\eit}{\end{itemize}}

\font\BFd=cmmib10
\font\BFt=cmmib10
\font\BFs=cmmib10 scaled 700
\font\BFss=cmmib10 scaled 500

\def\bbox#1{%
\relax\ifmmode
\mathchoice
{{\hbox{\BFd #1}}}
{{\hbox{\BFt #1}}}
{{\hbox{\BFs #1}}}
{{\hbox{\BFss #1}}}
\else \mbox{#1} \fi }

\def\k{{\bbox{k}}}

\def\u{{\bbox{u}}}

\newcommand{\MHz}{\mbox{MHz}}

\begin{document}



 
\submitted{\today. To be submitted to \apj.} 

\title{Detecting the rise and fall of 21 cm fluctuations with the Murchison Widefield Array}
\author{Adam Lidz\altaffilmark{1}, Oliver Zahn\altaffilmark{1,2}, Matthew McQuinn\altaffilmark{1},
Matias Zaldarriaga\altaffilmark{1,3}, Lars Hernquist\altaffilmark{1}}
\altaffiltext{1}{Harvard-Smithsonian Center for Astrophysics, 60 Garden Street,
Cambridge, MA 02138, USA}
\altaffiltext{2}{Berkeley Center for Cosmological Physics, Department of Physics, University of California, and Lawrence Berkeley National Labs, 1 Cyclotron Road; Berkeley, CA 94720, USA}
\altaffiltext{3}{Jefferson Laboratory of Physics; Harvard University; Cambridge, MA 02138, USA}
\email{alidz@cfa.harvard.edu}

\begin{abstract}

We forecast the sensitivity with which the Murchison Widefield Array
(MWA) can measure the 21 cm power spectrum of cosmic hydrogen,
using radiative transfer simulations to model reionization and the 21 cm
signal. The MWA is sensitive to roughly a decade in scale (wavenumbers
of $k \sim 0.1 - 1 h$ Mpc$^{-1}$), with foreground contamination
precluding measurements on larger scales, and thermal detector noise
limiting the small scale sensitivity. This amounts primarily to
constraints on two numbers: the amplitude and slope of the 21 cm power
spectrum on the scales probed.  We find, however, that the redshift
evolution in these quantities can yield important information about
reionization.  We
examine a range of theoretical models, spanning plausible
uncertainties in the nature of the ionizing sources, and the abundance
of gas-rich mini-halos during reionization. Although the 21 cm power spectrum
differs substantially among these models, a generic prediction is
that the amplitude of the 21 cm power spectrum on MWA scales ($k \sim
0.4 h$ Mpc$^{-1}$) peaks near the epoch when the 
intergalactic medium (IGM) is $\approx 50\%$
ionized.  Moreover, the slope of the 21 cm power spectrum on MWA
scales flattens as the ionization fraction increases and the sizes 
of the HII regions grow.
Considering detection sensitivity, we show that the optimal MWA
antenna configuration for power spectrum
measurements would pack all $500$ antenna tiles as close as
possible in a compact core. Provided reionization occurs in the MWA
observing band, this instrument
is sensitive enough in its optimal configuration to
measure redshift evolution in the slope and amplitude of the 21 cm
power spectrum. Detecting the characteristic redshift evolution of our
models will help confirm that observed 21 cm fluctuations originate
from the IGM, and not from foregrounds, and will provide an
indirect constraint on the evolution of the volume-filling factor of
HII regions during reionization. After two years of observations 
under favorable 
conditions, the MWA can constrain
the filling factor at an epoch when $\avg{x_i} \sim 0.5$ to within 
roughly $\pm \delta \avg{x_i} \sim 0.1$ at $2-\sigma$ confidence.
It can also constrain models for the ionizing sources and the abundance
of mini-halos during reionization.
\end{abstract}

\keywords{cosmology: theory -- intergalactic medium -- reionization -- large scale
structure of universe}

\section{Introduction} \label{sec:intro}

Detecting 21 cm emission from the high redshift IGM will provide fully
three-dimensional information on the epoch of reionization (EoR).
Futuristic experiments like the Square Kilometer Array (SKA) will have
the sensitivity to produce maps of reionization as a function of
redshift (Zaldarriaga et al. 2004, McQuinn et al. 2006). These maps,
taken in various frequency bands and with sufficiently high angular
resolution, will amount to a reionization `movie': they will depict
the growth of HII regions around individual sources and their
subsequent mergers with neighboring HII regions, and detail the
completion of the reionization epoch, whereby the entire volume of the
Universe becomes filled with ionized hydrogen.

While producing a reionization movie is perhaps the ultimate goal of
21 cm studies, first generation surveys like the MWA\footnote{The
array formerly known as the Mileura Widefield Array, is now known as
the Murchison Widefield Array, owing to a change of site.}  will
already provide valuable insights into the reionization process. These
first generation surveys will lack the sensitivity required to produce
detailed maps, but will allow for a statistical detection.  For
example, the MWA is expected to measure the power spectrum of 21 cm brightness
temperature fluctuations over roughly a decade in scale, in each of
several redshift bins (McQuinn et al. 2006, Bowman et al. 2006).

In this paper, we use radiative transfer simulations (Sokasian et
al. 2001, 2003, Zahn et al. 2007, McQuinn et al. 2007a, 2007b) to
quantify how well power spectrum measurements with the MWA can
constrain the reionization process. How well can we deduce the gross
features of the reionization movie from a statistical detection?  In
particular, we aim to check whether the MWA can constrain the volume
filling factor of HII regions as a function of time.  Reliable
estimates of the volume filling factor of HII regions -- the
`ionization fraction' of the IGM -- will pinpoint the timing and
duration of reionization.  This information, marking a key event in
the formation of structure in our universe, will constrain models for
the ionizing sources, and solidify existing measurements from cosmic
microwave background (CMB) polarization (Page et al. 2007), quasar
spectra (Fan et al. 2006, Lidz et al. 2006, 2007a, Becker et al. 2007,
Bolton \& Haehnelt 2007a, 2007b, Mesinger \& Haiman 2007, Wyithe et
al. 2005, Wyithe et al. 2007, Gallerani et al. 2007), Ly-$\alpha$
emitter (LAE) surveys (Malhotra \& Rhoads 2005, Kashikawa et al. 2006,
Furlanetto et al. 2006c, Dijkstra et al. 2007, McQuinn et al. 2007b,
Mesinger \& Furlanetto 2007a), and gamma-ray burst (GRB) optical
afterglow spectra (Totani et al. 2006, McQuinn et al. 2007c.), which
currently provide tantalizing, yet controversial and subtle to
interpret, clues.

The outline of this paper is as follows. In \S \ref{sec:rt_sig} we
describe our radiative transfer simulations, and examine the simulated
21 cm power spectrum and its redshift evolution. In \S
\ref{sec:pspec_noise} we forecast the sensitivity with which the MWA
can measure the 21 cm power spectrum, paying attention to foreground
contamination, and examining how the sensitivity depends on the, as
yet unfinalized, configuration of the MWA's antenna tiles (\S
\ref{sec:array_config}).  In \S \ref{sec:amp_slope} we show that the
MWA sensitivity generally boils down to constraints on each of the
slope and amplitude of the 21 cm power spectrum at $k \sim 0.1 -1 h$
Mpc$^{-1}$.  We then quantify how well the MWA can constrain these two
numbers in different redshift bins. In \S \ref{sec:fill_hii}, we show
how constraints on the slope and amplitude of the 21 cm power spectrum
in several redshift bins translate into constraints on the redshift
evolution of the volume-filling factor of HII regions during
reionization. In \S \ref{sec:conclusions} we summarize our main
results and discuss future research directions.

Throughout we consider a $\Lambda$CDM cosmology parameterized by: $n_s
= 1$, $\sigma_8 = 0.8$, $\Omega_m = 0.27$, $\Omega_\Lambda = 0.73$,
$\Omega_b = 0.046$, and $h=0.7$, (all symbols have their usual
meanings), consistent with the results from Spergel et al. (2007).

\section{The Redshift Evolution of the 21 cm Power Spectrum}
\label{sec:rt_sig}

In this section, we examine the redshift evolution of the 21 cm power
spectrum using the radiative transfer simulations of McQuinn et
al. (2007a, 2007b).  We focus on the power spectrum since the MWA
is expected to have limited imaging sensitivity, but can still provide
a statistical detection of 21 cm fluctuations.
First, let us briefly define terms.  In the limit
that the spin temperature of the 21 cm transition, $T_S$, is globally
much larger than the CMB temperature, $T_{\rm CMB}$, and ignoring
peculiar velocities, the 21 cm brightness temperature relative to the
CMB at spatial position $\vec{x_1}$ is:
\beqa
\delta_T(\vec{x_1}) = T_0 \avg{x_H} [1 + \delta_{\rm x}(\vec{x_1})][1 + \delta_\rho(\vec{x_1})].
\label{eq:tbright}
\eeqa
Here, $T_0$ is the 21 cm brightness temperature, relative to the CMB
temperature, at redshift $z$ and frequency $\nu = 1420/(1+z)$ $\MHz$,
for a neutral gas element at the cosmic mean gas density; $T_0 = 28
\left[(1+z)/10\right]^{1/2}$ mK in our adopted cosmology (Zaldarriaga
et al. 2004). The quantity $\avg{x_H}$ is the volume-averaged neutral
fraction, $\delta_{\rm x}$ is the fractional fluctuation in neutral
fraction, while $\delta_\rho$ is the fractional gas density
fluctuation.  We frequently quote results in terms of the
volume-averaged ionization fraction, $\avg{x_i} = 1 - \avg{x_H}$,
rather than in terms of the neutral fraction.

In general, we measure the power spectrum of $\delta_T(\vec{x_1})/T_0$
and plot the dimensionless power spectrum of this dimensionless
quantity, $\Delta^2_{\rm 21}(k) = k^3 P_{\rm 21}(k)/2 \pi^2$ -- i.e,
we plot the variance per logarithmic interval in wavenumber of the
field $\delta_T(\vec{x_1})/T_0$. We ignore peculiar velocities
throughout since they have little impact during the bulk of the
reionization epoch when ionization fluctuations are large (McQuinn et
al. 2006, Mesinger \& Furlanetto 2007b). We consider only
spherically-averaged power spectra since the MWA has limited
sensitivity in the transverse direction (McQuinn et al. 2006).

\subsection{Radiative Transfer Simulations}
\label{sec:rt_sims}

Let us briefly describe the simulations used in our analysis, and our
fiducial model for the ionizing sources and reionization
history. Radiative transfer is post-processed on an evolved N-body
simulation using the code of McQuinn et al. (2007a), a refinement of
the Sokasian et al. (2001, 2003) code, which in turn uses the adaptive
ray-tracing scheme of Abel \& Wandelt (2002). Some other approaches to
large scale reionization simulations are described in Iliev et
al. (2006), Kohler et al. (2005), Trac \& Cen (2006) and Croft \&
Altay (2007).  The radiative transfer calculation is performed on top
of a $130$ Mpc/$h$, $1024^3$ particle dark matter simulation run with
an enhanced version of Gadget-2 (Springel 2005). The minimum resolved
halo in this simulation is $\sim 10^{10} M_\odot$, but smaller mass
halos down to the atomic cooling mass (Barkana \& Loeb 2001), $M_{\rm
cool} \sim 10^8 M_\odot$, are incorporated with the appropriate
statistical properties as in McQuinn et al. (2007a). Ionizing sources
are placed in simulated halos with simple prescriptions. In our
fiducial model, we assume that a source's ionizing luminosity is
proportional to its host halo mass. More specifically, our fiducial
model is similar to the `S1' simulation of McQuinn et al. (2007a),
except that our model here was run in a slightly different cosmology, and in a
larger ($130$ Mpc/$h$ rather than $65.6$ Mpc/$h$) simulation box. Additionally,
the radiative transfer calculations (and our subsequent power spectrum 
measurements) are
computed on a $512^3$ grid, rather than a $256^3$ grid.

As we show subsequently, the MWA's ability to detect the 21 cm signal
from reionization depends strongly on the timing and duration of this
process (McQuinn et al. 2006, Bowman et al. 2006). Presently, we are
unable to predict reliably when reionization occurs, and how long it
lasts. This depends on many poorly constrained factors such as the
efficiency and initial mass function of star formation at high
redshift, the clumpiness of the high redshift IGM, the escape fraction
of ionizing photons from host galaxies, the degree to which
photoionizing and supernova feedback suppress star formation in low
mass galaxies, and other uncertain physics (see, e.g., the review by
Furlanetto et al. 2006a).  In our fiducial model, reionization
commences around $z \sim 11.5$, at which point $\sim 2\%$ of the
volume of the IGM is ionized, with $15 \%$ of the volume ionized by $z
= 8.8$, $54 \%$ by $z=7.3$, and $95 \%$ by $z=6.8$. The reionization
epoch may easily occur at significantly different redshifts than in
this model and likely satisfy all existing constraints mentioned in
the introduction, i.e., from CMB polarization, quasar spectra, LAE
surveys, and the optical afterglow spectra of GRBs.

However, our predictions for the power spectrum of ionization
fluctuations at a {\em given ionization fraction} are more robust than
our predictions at a {\em given redshift}, which suffer from the large
uncertainties noted above.  In particular, McQuinn et al. (2007a)
demonstrated that the ionization power spectrum varies weakly with
redshift when considered in a given model at fixed ionization
fraction, a consequence of the fact that, in most plausible models,
the bias of the ionizing sources is a weak function of redshift
(Furlanetto et al. 2006b).  The 21 cm power spectrum depends somewhat
more strongly on redshift at fixed ionization fraction than the
ionization power spectrum, since the 21 cm field explicitly involves
the density field, which evolves in time. Nevertheless, this
dependence is relatively weak during most of the EoR when the
ionization fluctuations are large (\S \ref{sec:model_dep}), and one
can think of our predictions in a given model as roughly invariant
with redshift for a given ionization fraction. Moreover, we aim to
extract information regarding the filling factor of HII regions from
the observations, so it will be convenient to consider the signal as a
function of this quantity.  Of course, as we noted earlier, the
uncertain mapping between ionization fraction and redshift 
can strongly impact the {\em detectability} of the signal
(\S \ref{sec:pspec_noise}).

\subsection{Simulated Power Spectrum in our Fiducial Model}
\label{sec:pspec_fid}

\begin{figure}
\bc
\includegraphics[width=9.2cm]{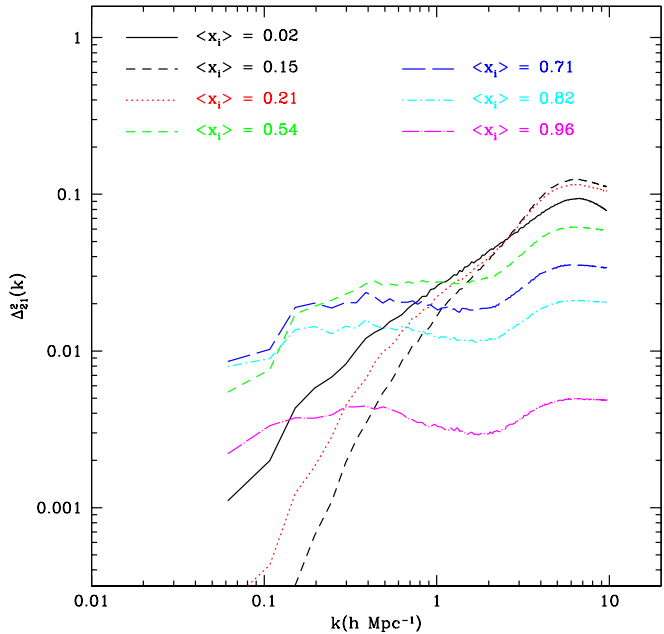}
\caption{Redshift evolution of the 21 cm power spectrum in our
fiducial model. The redshifts and volume-averaged ionization fractions 
shown are
$(\avg{x_i}, z) = $ (0.02, 11.46); (0.15, 8.76); (0.21, 8.34); (0.54,
7.32); (0.82, 6.90); and (0.96, 6.77).  The MWA probes wavenumbers
between roughly $k \sim 0.1-1 h$ Mpc$^{-1}$.  At the lowest ionization
fraction shown ($\avg{x_i} = 0.02$), the 21 cm power spectrum traces
the density power spectrum. There is a brief equilibration phase (see
text) where the 21 cm power spectrum amplitude drops on large scales
near $\avg{x_i} = 0.15$. The amplitude on MWA scales subsequently
increases until $\avg{x_i} \sim 0.5$, and then falls off at higher
ionization fractions. As the ionization fraction grows, after a brief
steepening during the equilibration phase, the slope of the power
spectrum on MWA scales flattens.  }
\label{fig:21cm_v_z}
\ec
\end{figure} 

In Figure \ref{fig:21cm_v_z}, we show the spherically-averaged 21 cm
power spectrum from our fiducial model for a range of redshifts and
ionization fractions. At early times and low ionization fraction ($z =
11.46$, $\avg{x_i} = 0.02$), the 21 cm power spectrum simply traces
the density power spectrum, except on very small scales where early
HII regions have some impact.  At slightly later times, there is a
brief phase (at $\avg{x_i} = 0.15, 0.21, z = 8.76, 8.34$) where the
large-scale 21 cm power spectrum falls below the density power
spectrum, and steepens in slope. This occurs because the large-scale
overdense regions initially contain more neutral hydrogen than
underdense ones, and consequently appear brighter in 21 cm.  On the
other hand, the overdense regions ionize first, and quickly transition
to being dimmer in 21 cm than underdense ones, which remain neutral.
This transition leads to a brief `equilibration' phase where overdense
and underdense regions have similar brightness temperatures, and the
large-scale 21 cm power spectrum is low as a result (Furlanetto et
al. 2004, Wyithe \& Morales 2007).  At these early stages of
reionization, however, our calculations may be inaccurate since we
neglect the impact of spin temperature fluctuations (Pritchard \&
Furlanetto 2007).

After the brief equilibration phase, the HII regions quickly grow,
boosting the large-scale power spectrum and suppressing the power on
small scales. As we demonstrate in the next section, the wavenumbers
relevant for the MWA are $k \sim 0.1-1 h$ Mpc$^{-1}$. On these scales,
the amplitude of the 21 cm power spectrum rises rapidly from when the
filling factor of ionized regions is $\avg{x_i} = 0.21$ to when
$\avg{x_i} = 0.54$, and then drops off at higher ionized fractions,
falling rather quickly when $\avg{x_i} \gtrsim 0.8$. In conjunction
with the increased power, the slope flattens and is close to $k^3
P_{\rm 21} (k) \propto const.$ for $\avg{x_i} \gtrsim 0.6$.

Provided that there are substantial ionization fluctuations on the
scales that the MWA is sensitive to, it is unsurprising that the 21 cm
power spectrum amplitude peaks around the epoch in which the IGM is
$\approx 50\%$ ionized. Recall that the variance in the ionization
field averaged on small scales, $\sigma_x^2 =\avg{x_i^2} -
\avg{x_i}^2$, must reduce to $\sigma_x^2 \sim \avg{x_i} - \avg{x_i}^2$
in the limit that each pixel is either completely neutral or
completely ionized. In this limit, the variance in the ionization
field peaks when $\avg{x_i} = 0.5$, close to the ionization fraction
at which the 21 cm power spectrum reaches its maximum on MWA scales.
Of course the 21 cm power spectrum is more complicated, since it
depends on the cross-correlation between ionization and overdensity
and higher order contributions (Lidz et al. 2007b), and since a large
portion of the ionization variance comes from scales that are not
probed by the MWA. Nevertheless, our simple argument motivates why the
maximum in fluctuation amplitude occurs near $\avg{x_i} \sim 0.5$.

In summary, although the MWA may be limited to measuring the power
spectrum over $\sim$ a decade in scale, our model 21 cm power spectra
evolve considerably with redshift and ionization fraction over this
range. Hence, sensitivity only to a decade in scale may still be quite
valuable, provided the MWA can make measurements over a number of
redshift intervals.

\subsection{Model Dependence}
\label{sec:model_dep}

\begin{figure}
\bc
\includegraphics[width=9.2cm]{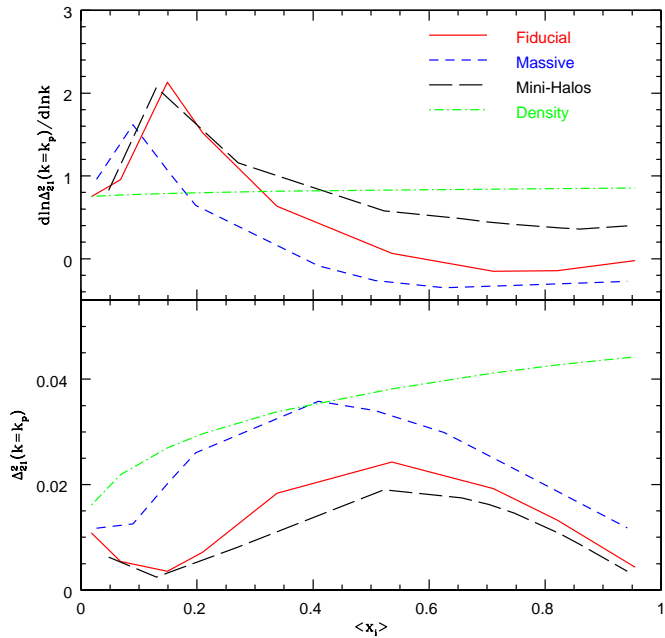}
\caption{Amplitude and slope of model 21 cm power spectra as a
function of ionization fraction. {\em Bottom}: Amplitude of the 21 cm
power spectrum, at the pivot wavenumber ($k_p=0.4 h$ Mpc$^{-1}$) for
MWA observations in our fiducial (solid red line), rare source (blue
short-dashed line), and mini-halo models (black long-dashed line),
plotted as a function of ionization fraction.  The green dot-dashed
line shows, for contrast, the amplitude of the density power spectrum
obtained by mapping redshift to ionization fraction as in our fiducial
model. {\em Top}: Slope of the 21 cm power spectrum at the pivot
wavenumber for our three models, as well as the density power spectrum
slope. The slight tilting of the slope of the density power spectrum
with decreasing redshift (increasing $\avg{x_i}$) owes to quasi-linear
effects.  The slope and amplitude of the 21 cm power spectrum vary
considerably among the different models at a given ionization
fraction. The behavior with ionization fraction across the different
models is, however, relatively generic: the amplitude of the 21 cm
power spectrum reaches a maximum close to the epoch when $\sim 50\%$
of the volume of the IGM is ionized, and the slope flattens with
increasing ionization fraction.  The maximum amplitude is reached at
slightly lower ionization fraction in our massive source model.
}
\label{fig:amp_v_x}
\ec
\end{figure}

How robust to model uncertainties is the rise and fall in 21 cm power
spectrum amplitude, and the flattening in power spectrum slope, at
increasing ionization fraction?  In order to check this, we examine
the amplitude and slope of the 21 cm power spectrum for two other
cases that roughly bracket model uncertainties.  McQuinn et
al. (2007a) investigated the many uncertain physical parameters that
can impact reionization and found that the nature of the ionizing
sources and the abundance of mini-halos have the strongest influence
on the ionization power spectrum.  If the sources are rare but very luminous
and lie predominantly in rather massive halos, then the high clustering of
the ionizing sources leads to larger HII regions at a given ionization 
fraction, and the ionization power spectrum is peaked on larger scales than in
our fiducial model. On the other hand, if mini-halos are abundant, the
HII regions are smaller, and the ionization power spectrum is peaked
on smaller scales than in our fiducial model (Furlanetto \& Oh 2005,
McQuinn et al. 2007a). The cross-correlation between ionization and
overdensity is also less significant with rare, efficient sources,
which acts in conjunction with the increased ionization power in these
models to boost the 21 cm power spectrum relative to our fiducial
model (Lidz et al. 2007b).

As a representative case with rare sources and large bubbles, we use a
model similar to `S3' of McQuinn et al. (2007a), except run in our
present larger volume.  Briefly, the S3 model has source luminosity $L
\propto M^{5/3}$: hence, the most massive, highly clustered halos
produce most of the ionizing photons.  Our representative case with
abundant mini-halos and small bubbles, adds mini-halos of dark matter
halo mass $M > 10^5 M_\odot$ into the simulation with the appropriate
statistical properties (McQuinn et al. 2007a). Each mini-halo is given
a cross section to ionizing photons initially equal to its halo virial
radius and we subsequently assume that each mini-halo is
photo-evaporated on a sound-crossing time.  This is a relatively
extreme situation since in reality the mini-halo cross sections will
shrink with time before they become completely photo-evaporated
(Shapiro et al. 2004, Iliev et al. 2005), and since many mini-halos
may be photo-evaporated by pre-heating prior to reionization (Oh \&
Haiman 2003).

We then determine the amplitude and slope of the 21 cm power spectrum
in each model, and plot these quantities as a function of ionization
fraction. The results of this calculation are shown in Figure
\ref{fig:amp_v_x}.  As a convenient choice, we consider the amplitude
and slope of the 21 cm power spectrum at a wavenumber $k_p = 0.4 h$
Mpc$^{-1}$, which is close to the middle of the range of wavenumbers
probed by the MWA (see \S \ref{sec:amp_slope}). We refer to call this
as the `pivot wavenumber' since we center our power law fits
in \S \ref{sec:amp_slope} on this value.

Concentrating first on the bottom panel of Figure \ref{fig:amp_v_x},
it is clear that the 21 cm power spectrum amplitude at fixed
ionization fraction differs significantly among the different
cases. The rare source model has the largest amplitude fluctuations,
the mini-halo model the smallest ones, and our fiducial model has an
intermediate level of fluctuations -- precisely the trends anticipated
above.  Note that, despite the presence of ionization fluctuations,
the amplitude of the 21 cm power spectrum at the pivot wavenumber is
generally less than that of the density power spectrum in our
models. This is a consequence of the ionization-density cross
correlation and higher order terms (Lidz et al. 2007b). However, the
relative amplitudes of the 21 cm power spectrum and the density power
spectrum are somewhat sensitive to the choice of pivot wavenumber. The
density power spectrum falls off more rapidly than the 21 cm power
spectrum towards large scales, and hence the 21 cm power spectrum
becomes larger than the density power spectrum for smaller wavenumbers
(see Figure \ref{fig:21cm_v_z}). The basic trend of 21 cm amplitude
with redshift or ionization fraction, however, is similar for most
wavenumbers within the MWA band.

Although the signal differs significantly between models, the 21 cm
power spectrum amplitude rises and falls with increasing ionization
fraction in each case.  The slight dip in power spectrum amplitude
near $15\%$ ionized fraction owes to the equilibration phase discussed
earlier.  Each model reaches a maximum amplitude rather close to an
ionization fraction of $\sim 50\%$, although the maximum in the
massive source model occurs at a slightly lower ionization fraction,
$\sim 40\%$.  The basic trend is very encouraging: the redshift at
which the 21 cm power spectrum reaches its maximum may provide {\em an
observational signature of the redshift at which the IGM is $\sim
50\%$ ionized.}  The ionization fraction at which the 21 cm power
spectrum amplitude reaches a maximum depends somewhat on our choice of
pivot wavenumber (see Figure \ref{fig:21cm_v_z}), although results are
similar near the middle of the range of wavenumbers probed by the MWA.
\footnote{Wyithe \& Morales (2007) find that the variance of the 21 cm
field, smoothed with a cylindrical top-hat of $10'$ angular scale,
reaches a {\em minimum} around $\avg{x_i} \sim 0.5$.  This angular
scale corresponds to $R \sim 19$ Mpc/$h$ and $k \sim 1/R \sim 0.05 h$
Mpc$^{-1}$ at $z=8$ in our cosmology, (although there is some ambiguity in
the proportionality factor relating $k$ and $1/R$).  Their 
cylindrical filter will
pass fluctuations on still larger scales, and so this is not
necessarily in contradiction with our results which find that the
power spectrum on somewhat smaller scales is {\em maximal} near
$\avg{x_i} = 0.5$. The disadvantage of their approach is that
foreground cleaning will prohibit measuring the large-scale modes
passed by their cylindrical filter. Their calculation methodology is
also accurate on scales only much larger then the sizes of individual
HII regions, which may impact their predictions during the middle
phase of reionization when HII regions are already quite large.}

Let us now turn our attention to our model predictions for the slope
of the 21 cm power spectrum on MWA scales, shown in the top panel of
Figure \ref{fig:amp_v_x}. In our fiducial model, the 21 cm power
spectrum is rather flat with $k^3 P(k) \sim const.$ at the end of
reionization. This is in contrast to the density power spectrum, which
has a slope of $k^3 P(k) \sim k^{0.8}$ near our pivot wavenumber.  A
similar flattening is seen in the massive source case, but here the
flattening occurs at smaller ionization fraction since this model
produces large ionized regions even at rather low ionization
fractions.  The slope of the power spectrum even goes slightly
negative after $\gtrsim 40\%$ of the volume is ionized.  On the other
hand, in our mini-halo model, the ionized regions are smaller at a
given ionization fraction, and there is consequently less flattening
with increasing ionization fraction.  The slight steepening seen at
early times in our models, near $\avg{x_i} \sim 0.15$, again owes to
the equilibration phase discussed in the previous section.

If the MWA finds that the slope of the power spectrum is significantly
flatter than that of the density power spectrum, this immediately
argues for the presence of large ionized regions, and implies that
reionization is well underway, while detecting fluctuations at all
clearly implies that less than $100\%$ of the IGM volume is
ionized.\footnote{Recently, Wyithe \& Loeb (2007) pointed out that
neutral gas in damped Ly-$\alpha$ (DLA) absorbers will produce a 21 cm
signal even after essentially the entire volume of the IGM is
reionized.  This signal will likely become comparable in amplitude to
that from the diffuse IGM only at the very end of reionization and can
potentially be distinguished from the diffuse IGM on the basis of 
its power spectrum shape. Specifically, we expect the DLA
contribution to the 21 cm power spectrum to have the form 
$P_{\rm 21, DLA} (k) \sim x_{\rm h, m}^2 \bar{b}^2 P_{\delta, \delta}(k)$,
where $x_{\rm h, m}$ is the mass-averaged neutral gas fraction locked up
in DLAs, $\bar{b}$ is the mass-averaged DLA bias, and
$P_{\delta, \delta}(k)$ is the matter power spectrum. For plausible numbers
of $x_{\rm h, m} = 0.05$, and $\bar{b} = 3$, the DLA contribution is a factor
of $5$ smaller than the diffuse IGM contribution in our fiducial model 
at $\avg{x_i} = 0.96$ and
$k_p \sim 0.4 h$ Mpc$^{-1}$. The diffuse component is even more dominant at
large scales, and less dominant on smaller scales. Hence, we expect the
DLA contribution to kick in essentially after reionization. The MWA
should therefore be able to study each of these interesting signals
separately.}  In the
mini-halo model, there is less steepening, and hence detecting a
relatively steep 21 cm power spectrum slope by itself does not imply
that one is probing an early phase of reionization, as it would for
our other cases.  It is clear from Figure \ref{fig:amp_v_x}, however,
that combining measurements of the 21 cm power spectrum amplitude and
slope in a few conveniently placed redshift bins will help move beyond
a mere detection of 21 cm fluctuations and allow constraints, albeit
indirect ones, to be placed on the ionization fraction as a function
of time.

Moreover, given the observational challenges anticipated for 21 cm
observations and the possibility that undesirable residuals survive
the foreground cleaning process, simple theoretical diagnostics such
as the ones mentioned here are extremely valuable.  Observing the
power spectrum amplitude rise and fall with redshift, and the power
spectrum slope flatten, is characteristic of the anticipated
reionization signal, but foreground residuals are unlikely to mimic
this behavior.

\section{MWA Power Spectrum Sensitivity}
\label{sec:pspec_noise}

We now consider the statistical significance at which the MWA might
detect this characteristic redshift evolution in the 21 cm power
spectrum.  We first write down the equations describing statistical
error estimates for the 21 cm power spectrum (Zaldarriaga et al. 2004,
Morales 2005, McQuinn et al. 2006). We generally follow the notation
of Furlanetto \& Lidz (2007).

The variance of a 21 cm power spectrum estimate for a single $\k$-mode
with line of sight component $k_\parallel = \mu k$, restricting
ourselves to modes in the upper-half plane, considering both sample
variance and thermal detector noise, and assuming Gaussian statistics,
is given by:
\beqa
\sigma_P^2(k,\mu) = \left[P_{\rm 21}(k, \mu) + \frac{T^2_{\rm sys}}{T_0^2}\frac{1}{B t_{\rm int}}
\frac{D^2 \Delta D}{n(k_\perp)}\left(\frac{\lambda^2}{A_e}\right)^2\right]^2.\nonumber \\
\label{eq:var21}
\eeqa
The first term in Equation (\ref{eq:var21}) is the contribution 
from sample variance, while the second term comes from thermal
noise in the radio telescope. The thermal noise term depends on the
system temperature, $T_{\rm sys}$, the co-moving distance to the
center of the survey at redshift $z$, $D(z)$, the survey depth,
$\Delta D$, the observed wavelength, $\lambda$, the effective area of
each antenna tile, $A_e$, the survey bandwidth, $B$, the total
observing time, $t_{\rm int}$, and the distribution of antennae. The
dependence on antenna configuration is encoded in $n(k_\perp)$ which
denotes the number density of baselines observing a mode with
transverse wavenumber $k_\perp$ (McQuinn et al. 2006). The factor of
$T_0^2$ in the noise term is appropriate because we consider the error
in the power spectrum of $\delta_T/T_0$ (Equation \ref{eq:tbright}).

In order to estimate the variance of the power spectrum averaged over
a spherical shell of logarithmic width $\epsilon = d \rm{ln}k$, we add
the statistical error for individual $k$-mode estimates in inverse
quadrature:
\beqa
\frac{1}{\sigma_P^2(k)} = \sum_\mu \frac{\epsilon k^3 V_{\rm survey}}{4 \pi^2} 
\frac{1}{\sigma_P^2(k,\mu)}.
\label{eq:var_shell}
\eeqa
In this equation, $V_{\rm survey} = D^2 \Delta D
\left(\lambda^2/A_e\right)$ denotes the effective survey volume of our
radio telescope, and the sum is over all modes in the upper half plane
contained within the survey volume. The maximum $\mu$ included in the
sum over modes is set by the survey depth, up to a maximum possible of
$\mu=1$, while the minimum $\mu$ depends on the highest transverse
wavenumber sampled by the array, down to a minimum possible of
$\mu=0$. In practice, we approximate this sum by an integral.  We
generally plot the error in $\Delta^2_{\rm 21}(k)$ which is related to
the above $\sigma_P(k)$ by a factor of $k^3/2\pi^2$.

Making the array more compact increases the number density of
baselines, $n(k_\perp)$, sampling low $k_\perp$, but truncates the sum
over modes (Equation \ref{eq:var_shell}) at larger $\mu_{\rm min}$. An
important point to note is that, owing to high-$k$ line of sight modes
in the sum of Equation (\ref{eq:var_shell}), the MWA can still
estimate the power spectrum for wavenumbers with $k > k_{\rm \perp,
max}$, where $k_{\rm \perp, max}$ is the highest transverse wavenumber
probed by the antenna array. Indeed, for the rather compact array
configurations expected for the MWA, the sensitivity is concentrated
along the line of sight, and the sum over $\mu$ in Equation
(\ref{eq:var_shell}) is restricted to a limited range close to $\mu=1$
for high $k$ modes.

\subsection{Assumed MWA Survey Parameters}
\label{sec:mwa_specs}

The MWA will be an array of $N_a = 500$ antenna tiles observing a wide
field on the sky of $\sim 800$ deg$^2$ at frequencies of $80-300$
$\MHz$, corresponding to 21 cm emission redshifts of $z=4-17$. Each
antenna tile consists of $16$ dual polarization dipoles layed out in a
$4$ m-by-$4$ m grid, and covers an effective collecting area of $A_e =
14$ m$^2$ at $z=8$ (Bowman et al. 2006).  We linearly interpolate
between the values of effective area given in Table 2 of Bowman et al.
(2006) to determine the effective antenna areas at other redshifts.
We assume that the system temperature is set by the sky temperature,
which we take to be $T_{\rm sys} = 280 ((1+z)/7.5)^{2.3}$ K, following
Wyithe \& Morales (2007). We consider observations over a bandwidth of
$B = 6$ $\MHz$, and a total observing time of $t_{\rm int} = 1,000$
hrs.  This integration time represents an optimistic estimate for the
total amount of usable observing time the MWA might achieve within a
single calendar year.  The bandwidth is chosen to be small enough to
ensure that the signal evolves minimally over the corresponding
redshift interval (McQuinn et al. 2006), which is only 
$\Delta z = 0.3$ at $z=8$.

We will consider two choices for the antenna distribution, since the
sensitivity of the array depends strongly on the configuration of the
tiles (McQuinn et al. 2006, Bowman et al. 2006, \S
\ref{sec:array_config}).  In the first configuration, following Bowman
et al. (2006), we assume that a fraction of
the $500$ antenna tiles are packed as close as physically possible
within a $20$ m core, and that the remaining antennae follow an
$r^{-2}$ distribution out to a maximum baseline of $1.5$ km. This
configuration has $\sim 80$ antennae in the compact $20$ m core.  In
addition, we consider a configuration where {\em all} $500$ antennae
are packed as close as physically possible within a $\sim 50$ m core,
which we will term the `super-core' distribution. For each
configuration, we calculate the approximate number density of
baselines observing a given transverse mode, $n(k_\perp)$, by
determining the auto-correlation function of the antenna
distribution. In practice, one may not be able to place the antennae
as close as physically possible, since, e.g., the tiles will interfere
with one another when packed too tightly. Nevertheless, our examples
represent useful models to understand how sensitivity depends on array
configuration.

\subsection{Results}
\label{sec:sens_results}

\begin{figure}
\bc
\includegraphics[width=9.2cm]{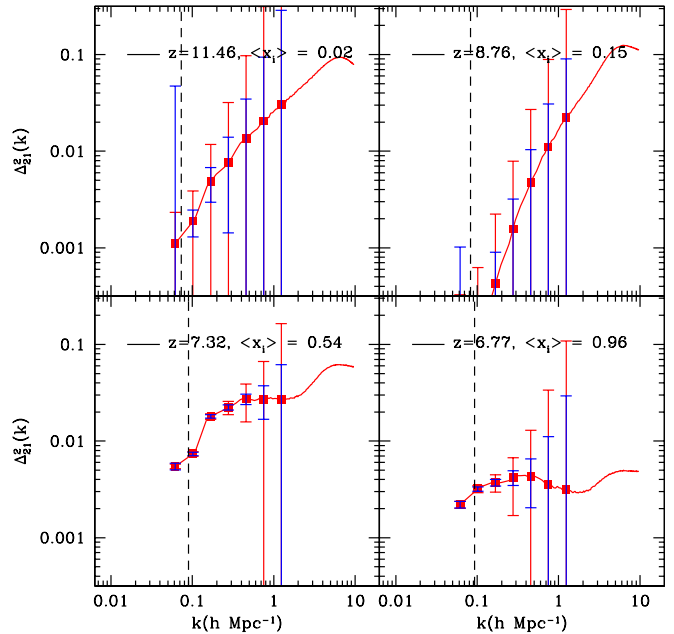}
\caption{21 cm power spectrum sensitivity for the MWA at different redshifts for our
fiducial model. The red curves show the theoretical model 21 cm power spectra from our
fiducial model (from Figure \ref{fig:21cm_v_z}) at different redshifts and ionization
fractions, as labeled. The red error bars indicate statistical error estimates for the
MWA with $1,000$ hours of observation, and an $r^{-2}$ antenna distribution.
The smaller blue error bars show the expected statistical error bars for 
the MWA with 
antennae arranged in the super-core configuration (see text). 
Foreground cleaning over our assumed $6$ $\MHz$ bandwidth prohibits power spectrum
measurements on wavenumbers smaller than indicated by the black dashed lines.
}
\label{fig:pspec_err_v_z}
\ec
\end{figure}

We estimate the statistical error bars from our fiducial model 21 cm
power spectrum calculations (Figure \ref{fig:21cm_v_z}), using the MWA
survey specifications, and Equations (\ref{eq:var21}) and
(\ref{eq:var_shell}), and considering spherical shells of logarithmic
width $\epsilon = 0.5$. Note that our hypothetical surveys extend to
much higher wavenumbers along than transverse to the line of sight,
and so our surveys do not sample full spherical shells in $k$-space at
high wavenumber -- they only sample a limited fraction of a $k$-space
shell close to the line of sight axis.

In this section, our calculations are similar to previous work
(McQuinn et al. 2006, Bowman et al. 2006), except here we consider a
more realistic model for the 21 cm power spectrum.  The results of
these calculations are shown in Figure \ref{fig:pspec_err_v_z} for
four example redshifts and ionized fractions.  The dashed lines in the
figure indicate wavenumbers below which foreground removal -- for our
particular choice of bandwidth, $B = 6$ $\MHz$ -- will prohibit power
spectrum measurements. The foreground cleaning process exploits the
expectation that the foregrounds are spectrally smooth, while the
signal has structure in frequency space (e.g. Zaldarriaga et al. 2004,
Morales \& Hewitt 2004, Morales et al. 2006, McQuinn et al. 2006). This process 
will -- at
the very minimum -- remove all line of sight modes with $k_\parallel
\leq 2 \pi/\Delta D$. The discreteness of modes in the survey then
implies that {\em all} modes with $k \leq 2 \pi/\Delta D$ will be lost
in the foreground cleaning process.  
At high wavenumber, the antenna array samples modes poorly owing to its
limited angular resolution, and hence power measurements at wavenumber
above $k \gtrsim 1 h$ Mpc$^{-1}$ are quire noisy.

Measurements early in the EoR generally have low signal to noise. This
is illustrated in the top two panels of Figure
\ref{fig:pspec_err_v_z}. The sensitivity of the measurements degrades
significantly towards high redshift because of the increasing sky
brightness towards low frequency, $T_{\rm sys} \propto (1+z)^{2.3}$,
and because the signal is weaker on large scales at early times. In
this particular model, the $z = 8.76$, $\avg{x_i} = 0.15$ measurement
actually has less signal to noise than the higher redshift $z =
11.46$, $\avg{x_i} = 0.02$ model. This is because the $z=8.76$ output
is near the equilibration phase (see \S \ref{sec:pspec_fid}), where
the large scale power actually drops below the density power
spectrum. At lower redshifts, during the intermediate phase of
reionization, the MWA will make highly significant power spectrum
measurements over a range in scales, as illustrated by the lower left
panel where $z = 7.32, \avg{x_i} = 0.54$.  Indeed, even when the IGM
is $96 \%$ ionized at $z=6.77$ in this model, 21 cm fluctuations
should be detectable with the MWA in its $r^{-2}$ arrangement, with a
strong detection expected for the super-core configuration.

To provide a quantitative measure, we calculate the total $S/N$ at which
the MWA can measure the 21 cm power spectrum. We compute the total
$S/N$ by taking the square root of the sum of the squares of the $S/N$
in each $k$-bin. For our $r^{-2}$ antenna configuration, this
calculation gives $\sim 1-\sigma, 0.5-\sigma, 24-\sigma$, and
$13-\sigma$ at $(\avg{x_i}, z) = (0.02, 11.46); (0.15, 8.76); (0.54,
7.32)$ and $(0.96, 6.77)$ respectively. In the super-core
configuration, these numbers are boosted to $\sim 4-\sigma, 2-\sigma,
39-\sigma$ and $25-\sigma$ respectively. 
Taken at face value, these numbers imply that the MWA, in its super-core
configuration, can achieve
a $3-\sigma$ power spectrum detection
when the IGM is $50\%$ neutral after only $t_{\rm int} \sim 100$ hrs of
integration time.
Note, however, that most of
the detection sensitivity comes from the first $k$-bin beyond the
foreground limit. Given that this first bin is potentially most
impacted by residual foregrounds (McQuinn et al. 2006), and given that
the discriminating power between models is slightly larger at higher
$k$, we also quote the $S/N$ at which one can measure power in the
$k$-bin with $k=0.46 h$ Mpc$^{-1}$ (and $\epsilon = d{\rm ln k} = 0.5$), near 
the middle of the range of
scales probed by the MWA. In the super-core configuration, the $S/N$
for detecting power in this $k$-bin is $0.6-\sigma, 0.8-\sigma,
8-\sigma$, and $2-\sigma$ for the respective $(\avg{x_i}, z)$ above
(see also \S \ref{sec:amp_slope}).

\begin{figure}
\bc
\includegraphics[width=9.2cm]{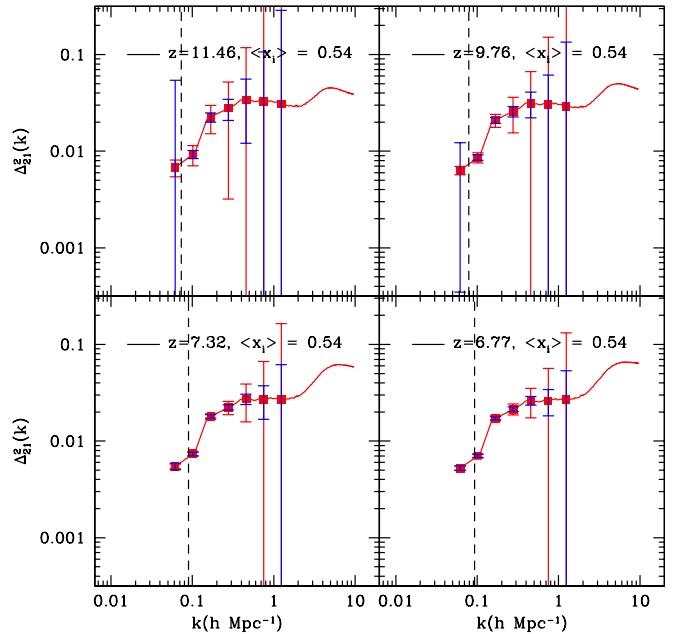}
\caption{21 cm power spectrum sensitivity for the MWA at different redshifts
with $\avg{x_i} \sim 0.5$ at each redshift. Similar to Figure \ref{fig:pspec_err_v_z} except here we show how the sensitivity depends on redshift for
a {\em fixed ionization fraction} of $\avg{x_i} \sim 0.5$.}
\label{fig:variance_x_0.5_v_z}
\ec
\end{figure}

To get a sense for how the power spectrum signal to noise depends on
{\em when} reionization occurs, we repeat our sensitivity calculations
for models in which $\avg{x_i} = 0.54$ at each of $z=6.77,$ $9.76$ and
$z=11.46$.  We contrast these models with our earlier calculations in
which $\avg{x_i} = 0.96, 0.07, 0.02$ at these respective redshifts.
In order to avoid running additional radiative transfer calculations,
we construct 21 cm fields at these redshifts using in each case the
ionization field in our fiducial model at $z=7.32$ (which has
$\avg{x_i} = 0.54$), in conjunction with the simulated density field
at each desired redshift. This should be a good approximation since
the ionization power in a given model depends mainly on the ionization
fraction and not explicitly on redshift (McQuinn et al. 2007a).

The results of this calculation are shown in Figure
\ref{fig:variance_x_0.5_v_z}. Comparing with the results of Figure
\ref{fig:pspec_err_v_z}, we see that the 21 cm power spectrum
sensitivity at a given redshift is enhanced significantly when
$\avg{x_i} \sim 0.5$ compared to the sensitivity at the beginning or
end of reionization. The large ionized bubbles during the middle phase
of reionization significantly boost the amplitude of the 21 cm power
spectrum on large scales, and facilitate detection above the thermal
detector noise.  Quantitatively, the total $S/N$ at which the MWA can
detect the 21 cm power spectrum at $z=11.46$ in our fiducial model (in
which case $\avg{x_i} = 0.02$) is only $1-\sigma$ for the $r^{-2}$
antenna configuration, and $4-\sigma$ for the super-core distribution.
If $\avg{x_i} \sim 0.5$ at this redshift, however, the $S/N$ increases
to $5-\sigma$ in the $r^{-2}$ configuration, and $14-\sigma$ for the
super-core arrangement.  For comparison, the $S/N$ at which the MWA
can detect power in the $k=0.46 h$ Mpc$^{-1}$ bin is $1.5-\sigma,
3.4-\sigma$ and $9.9-\sigma$ when $\avg{x_i} = 0.54$ at $z=11.46,$
$9.76$, and $6.77$ respectively. Interestingly, despite the
increased sky noise towards high redshift, the MWA can still expect a
detection at $z \sim 11$ if $\avg{x_i} \sim 0.5$ at this redshift.

It is also interesting to examine how strong a lower limit on the ionization
fraction the MWA might place from a null detection. If the IGM
is mostly ionized at $z=8$, for example, should we expect the MWA to detect 21 cm power at this redshift?
To investigate this, we use the ionization fields from our fiducial model
at $\avg{x_i} = 0.82, 0.96$ to produce mostly ionized 21 cm power spectrum
models at $z=8.15$.
Interestingly, with our fiducial survey parameters these
models both yield $\gg 3-\sigma$ power spectrum detections in
each of the $r^{-2}$ and super-core configurations, provided
we include all modes with wavenumber larger than our foreground cut. For
comparison, the $S/N$ for detecting power in the
$k=0.46 h$ Mpc$^{-1}$ bin alone is larger than $3-\sigma$ only for the 
$\avg{x_i} = 0.82$ model in the super-core configuration. The super-core
configuration produces only a $1-\sigma$ power spectrum detection in this 
bin at
$\avg{x_i} = 0.96$, $z=8.15$, while neither $\avg{x_i}$ model produces a 
significant
detection in this bin with the $r^{-2}$ antenna configuration. 
Hence if the MWA fails to detect power at $z \leq 8$, this will
still provide a stringent lower limit on $\avg{x_i}$. Since much of the power
spectrum sensitivity comes from $k$-bins close to the foreground limit, however,
the precise lower limit depends on how well foreground cleaning algorithms
perform for wavenumbers close to the foreground cut (McQuinn et al. 2006). This
is an important topic for further study.

\subsection{Sensitivity to Array Configuration}
\label{sec:array_config}

As noted earlier (Zaldarriaga 2004, McQuinn et al. 2006, Bowman et
al. 2006), the sensitivity of 21 cm power spectrum measurements
depends strongly on the array's antenna tile configuration. Here we
expand on this point, advocating a still more compact antenna
distribution than previous authors.

Concentrating on the fiducial model output at $z=7.32$ where the $S/N$
of the detection is highest, it is clear that the super-core
configuration yields significantly smaller statistical error bars.  At
$k = 0.1 h$ Mpc$^{-1}$ the statistical error bars are a factor of
$1.3$ smaller in the super-core configuration than for the $r^{-2}$
distribution, but by $k = 1 h$ Mpc$^{-1}$, the error bars are a factor
of $\sim 4$ smaller in the super-core configuration.  The gain in
sensitivity at low $k$ is less substantial because the large scale
modes become sample variance dominated, and hence making the array
more compact beyond some point does not further boost power spectrum
sensitivity. Interestingly, the fact that large scale modes become
sample variance dominated in the super-core configuration, implies
that the MWA is capable of imaging large scales modes in this
arrangement -- i.e., the signal to noise per mode is unity for some
large scale modes. We will consider the MWA's imaging capabilities in
future work. Here we focus on the array's sensitivity for statistical
power spectrum measurements.

\begin{figure}
\bc
\includegraphics[width=9.2cm]{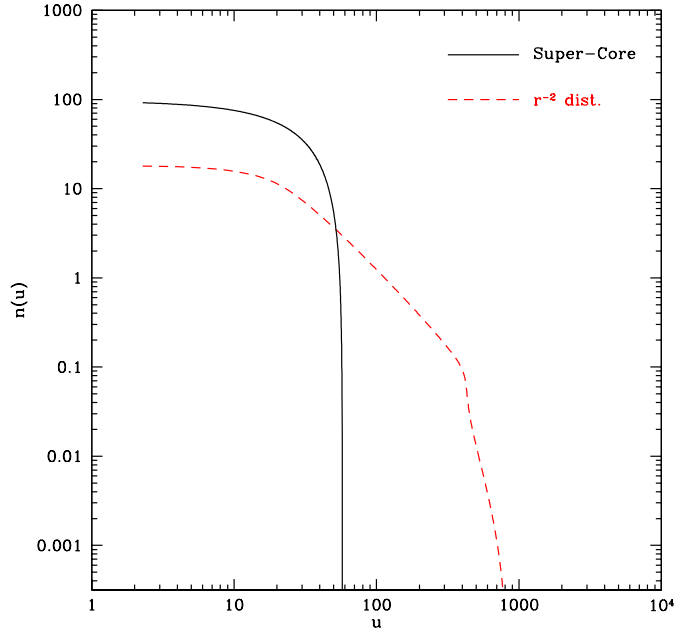}
\caption{Density of baselines observing a given mode (or visibility $u$) for our
two model antenna configurations at $z=7.32$. The density of baselines drops off rapidly
towards large visibilities for both arrangements. The super-core configuration
provides superior sampling of the small visibilities. 
}
\label{fig:dens_visib}
\ec
\end{figure}

The increased sensitivity of the array in its compact core
configuration arises because the long baselines in the $r^{-2}$
distribution are too sparsely sampled to be useful. To illustrate
this, we plot the number density of baselines observing a given
visibility (see Equation \ref{eq:var21}) for each antenna
configuration (see also Bowman et al. 2006).  The visibility $|\u|$ is
related to the transverse wavenumber by $2 \pi |\u| = k_\perp D(z)$.
Integrating $n(|\u|)$ over all $\u$ gives the total number of antenna
pairs in the array.

The $r^{-2}$ antenna distribution is relatively flat at low $u$ owing
to this configuration's $\sim 20 m$ core region where the antennae are
stacked as close as physically possible, while it falls off steeply
towards high $u$. Note that the thermal noise contribution to the
power spectrum variance scales as $\sigma_P(k, \mu) \propto
1/n(k_\perp)$ (Equation \ref{eq:var21}), and so the rapidly
diminishing number of antenna pairs towards high $u$ implies a
correspondingly rapid increase in the thermal noise.  Indeed, by $u =
85$ the number density of baselines falls by an order of magnitude
from its value at the smallest visibilities sampled, and the thermal
noise in these high $k_\perp$ modes becomes prohibitive.  In our
super-core configuration, the antennae are packed as close as
possible, doing away with the long baselines entirely, yet
significantly increasing the sensitivity to low $k_\perp$ modes.

Note that the limited resolution of our array configurations in the
transverse direction does not prohibit measuring high-$k$ modes
entirely, owing to the array's superior line of sight resolution
(Morales 2005, Bowman et al. 2006, McQuinn et al. 2006). Instead, our
sum over modes in Equation (\ref{eq:var_shell}) is simply restricted
to modes close to the line of sight, i.e. to modes with $\mu$ larger
than $\mu_{\rm min} = [1 - k_{\perp, max}^2/k^2]^{1/2}$. While this
prohibits probing the full angular structure of the 21 cm power
spectrum, this is not a big limitation during most of the EoR when
peculiar velocities have little impact and the power spectrum is
essentially isotropic.  Our calculations hence argue that the optimal
antenna configuration for power spectrum measurements places all $500$
antennae as close as physically possible in a compact core.

\subsection{Array Configuration Trade-offs}
\label{sec:array_tradeoffs}

\begin{figure}
\bc
\includegraphics[width=9.2cm]{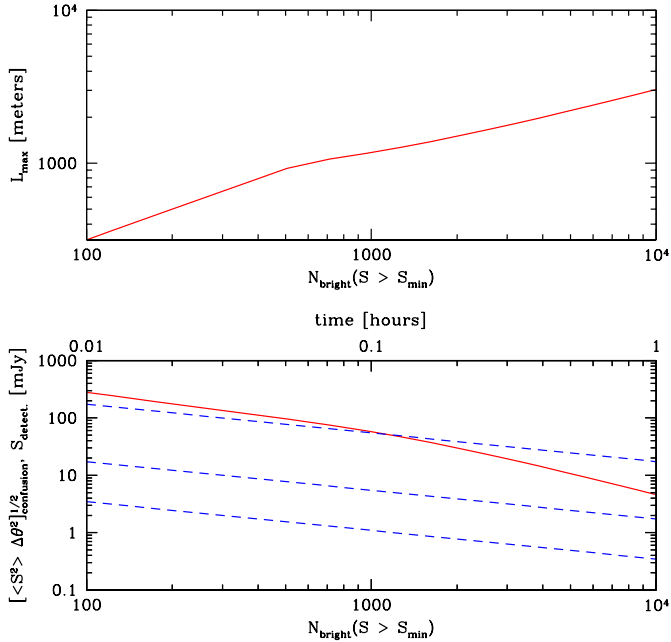}
\caption{Requirements for detecting bright point sources and calibrating
the array. {\em Top}: The angular resolution, expressed as a baseline length,
needed to detect $N_{\rm bright}$ sources above confusion noise from unresolved
sources at $10-\sigma$ significance. {\em Bottom}: The red solid line is the level
of confusion noise as a function of the number of bright sources needed for calibration
purposes. The resolution, $\Delta \theta$, is adjusted (as in the top panel) to allow
for a $10-\sigma$ detection of each point source. The blue dashed lines show the thermal
detector noise for $N_{\rm a, long} = 10, 100$, and $500$ antennae at long baselines (from
top to bottom), plotted as a function of observing time (top horizontal axis). Approximately
$50-100$ antennae at long ($\sim 1 km$) baselines suffice to detect a few hundred point
sources above the thermal and confusion noise, after integrations of $t_{\rm int} \sim$
tens of seconds.
}
\label{fig:calib}
\ec
\end{figure}

There are of course some tradeoffs involved in making the array more
compact. For one, in order to calibrate the array, high angular
resolution is needed to detect bright point sources above the
confusion noise from unresolved sources. Let us make a rough estimate
for how stringent these requirements are. 

We assume that in order to calibrate the array one needs to detect
$N_{\rm bright}$ sources above confusion and thermal detector noise
after integrating for a duration of $t_{\rm int}$ hours. In practice,
the MWA likely requires of order $N_{\rm bright} \sim$ a few hundred
bright sources for antenna calibration, and needs to be able to detect
these bright sources on time scales of $t_{\rm int} \sim$ tens of
seconds. The number of bright sources required for antenna calibration
depends on the total number of antennae, and the number of calibration
parameters per antenna. The calibration timescale is set by the
timescale over which ionospheric distortions vary. We leave these
quantities as free parameters in our calculation to show how
calibration requirements scale with the number of bright sources
required, and the calibration timescale.

For these calculations, we adopt the Di Matteo et al. (2002) model for
radio source counts.  This model is based on radio counts from the 6C
catalog at $151$ $\MHz$ (Hales et al. 1988), with an extrapolation to
low source flux. The differential source count, $dn/dS$, is given by
$dn/dS = 4 (S/S_0)^{-1.75}$ sources mJy$^{-1}$ str$^{-1}$, for $S <
S_0$, and $dn/dS = 4 (S/S_0)^{-2.51}$ sources mJy$^{-1}$ str$^{-1}$,
for $S > S_0$, with $S_0 = 880$ mJy.\footnote{The notation in Di
Matteo et al. (2002) is ambiguous, and their model for the
differential source counts has been misinterpreted by Gnedin \& Shaver
(2004).  The differential source count given above matches the bright
end measurements of Hales et al. (1988), and is continuous across
$S_0$ where the slope flattens.} The number of bright sources
detectable by the MWA above some limiting minimum flux, $S_{\rm min}$,
across the entire MWA field of view, $\Delta \Omega$, is: $N(> S_{\rm
min}) = \Delta \Omega \int_{S_{\rm min}}^{\infty} dS dn/dS$.  The
Poisson noise from unresolved sources below the detection limit is
\footnote{Using the model of Di Matteo et al. (2002), we find that
the variance owing to the clustering of unresolved radio sources is 
much smaller than the Poisson noise for the bright source cuts and
angular scales of interest. We hence neglect clustering here.} :
\beqa
\avg{S^2} = \int_{0}^{S_{\rm min}} dS S^2 dn/dS
\label{eq:poiss_unr}
\eeqa
The resulting confusion noise, given an MWA pixel of angular size
$\Delta \theta$, is $\sim \left[\avg{S^2} (\Delta \theta)^2
\right]^{1/2}$. In order to detect a point source of flux $S_{\rm
min}$ above the confusion noise from unresolved sources, we would like
the flux from this source to be say $10$ times as large as the
confusion noise, $S_{\rm min}/\left[\avg{S^2} (\Delta \theta)^2
\right]^{1/2} = 10$. This requirement demands high angular
resolution: one needs relatively good angular resolution in order to
beat down the confusion noise from the unresolved sources.

We investigate the angular resolution required to overcome confusion
noise from unresolved point sources in Figure \ref{fig:calib}. First,
we determine the minimum flux, $S_{\rm min}$, above which the MWA can
detect $N_{\rm bright}$ point sources in its entire field of view. We
then solve for the angular resolution $\Delta \theta$ such that the
confusion noise from point sources with $S < S_{\rm min}$ is less
than $0.1 S_{\rm min}$ -- i.e., we require that the source is
detectable above the confusion noise at $10-\sigma$ significance. The
angular resolution satisfying this constraint, $\Delta \theta$,
corresponds to a baseline with length $L_{\rm max} \sim \lambda_{\rm
obs}/\Delta \theta$. The figure illustrates that, in order to detect a
few hundred point sources above the confusion noise -- roughly the
number of bright point sources required for MWA array calibration --
one requires baselines of order $\sim 1$ km. Hence at least some
antennae are required beyond the super-core configuration, which
provides a maximum baseline of only $\sim 100$ m.

In order to determine {\em how many} antennae are needed at long
baselines, we want to demand that the thermal detector noise of the
large baseline antennae is much less than the source flux, $S_{\rm
min}$, that we aim to detect. Let us take $N_{\rm a, long}$ antennae
at long baselines. The detector noise for point source detection, with
$N_{\rm a, long}$ antennae, each of effective area $A_e$, integrating
for $t_{\rm int}$ hours over a bandwidth $B$ is:
\beqa
S_{\rm therm} = \frac{2 k_B T_{\rm sys}}{A_e N_{\rm a, long} \sqrt{2 t_{\rm int} B}} .
\label{eq:stherm}
\eeqa 
We plot the thermal noise in the bottom panel of Figure
\ref{fig:calib} for a bandwidth of $B = 6$ $\MHz$, and $A_e = 14$
m$^{2}$ (appropriate for the MWA at $z = 8$; Bowman et al. (2006)), as
a function of $t_{\rm int}$ for several different values of $N_{\rm a,
long}$. On timescales of $\sim$ tens of seconds, the thermal noise
expected for $\sim 100$ antennae at long baselines is much less than
both the confusion noise and the minimum source flux required to
detect a few hundred bright sources. This suggests that while some
antennae are required at large baselines, the requirements are rather
modest. In particular, only $50-100$ long baseline antennae are
required by our simple estimate, leaving $400-450$ antennae for our
super-core configuration. This is conservative, since we have effectively
treated the antennas in the core as a single antenna in considering
the calibration requirements.
Our proposed arrangement is in marked contrast to
the suggested $r^{-2}$ distribution, in which only $\sim 80$ antennae
are in a compact core.

From our simple estimates, it appears that the super-core, or a
similar configuration, is feasible since only a handful of long
baselines appear necessary for array calibration.  This should, of
course, be tested with detailed simulations of the MWA pipeline,
performed for a variety of antenna distributions.  While we have
argued that the super-core is optimal for spherically-averaged 21 cm
power spectrum estimates, it may not be optimal for other
programs. Heliospheric science, and surveys for radio transients for
example, will naturally require high angular resolution and favor
less compact antenna configurations.  Other EoR science projects may
be impacted as well. For example, a compact-core is not ideal for
detecting the 21 cm-galaxy cross power spectrum: here one needs to
balance the MWA's high sensitivity along the line of sight, but poor
transverse sensitivity, with the galaxy survey's poor line of sight
sensitivity -- owing to uncertainties in photometric redshifts -- yet
superior transverse sensitivity (Furlanetto \& Lidz 2007). The MWA
program to image quasar HII regions (Geil \& Wyithe 2007) may also
suffer from reduced angular resolution. In practice, the MWA might
start with an $r^{-2}$ configuration, or a slightly more compact
arrangement, and gradually add antennae into a compact core.

\section{Constraining the Power Spectrum Amplitude and Slope}
\label{sec:amp_slope}

Figure \ref{fig:pspec_err_v_z} suggests that the MWA will mainly be
sensitive to the amplitude and the slope of the 21 cm power spectrum
at each of several redshifts.  Constraints on just these two numbers,
in several redshift bins, are still quite interesting: we showed in \S
\ref{sec:model_dep} that the 21 cm power spectrum amplitude and slope
evolve with redshift in a relatively generic and informative
manner. Let us consider how well we can determine the parameters of a
power law fit to the MWA power spectrum measurements in several
redshift bins.

In practice, the MWA will be limited by large data rates to handling
$32$ $\MHz$ intervals of bandwidth at a time. This $32$ $\MHz$
interval can be sampled in any manner desired from the full spectral
range covered by the instrument, $80-300$ $\MHz$ -- e.g., one could
take $20$ $\MHz$ centered around $z \sim 6$ and $12$ $\MHz$ around $z
\sim 10$, or one could take a contiguous $32$ $\MHz$ stretch, etc. One
can further sub-divide this $32$ $\MHz$ into smaller intervals for
foreground-cleaning and power spectrum analyses. The best choice of
frequencies to analyze depends, of course, on when reionization
occurs.

For illustrative purposes, let us start with a $32$ contiguous $\MHz$
bin centered fortuitously on $z = 7.4$, when reionization is roughly
$50\%$ complete in our model.  We discuss less fortunate choices
subsequently.  We divide this up into five $6$ $\MHz$ redshift bins in
which we separately estimate power spectrum errors, and discard the
remaining $2$ $\MHz$ in our analysis. We consider observing this $32$
$\MHz$ stretch for a year, assuming, as above, that this yields
$1,000$ hours of usable data.  This stretch of $32$ $\MHz$ corresponds
to a redshift range of $z = 6.7 - 8.2$.  

In addition, we consider observations of a further contiguous stretch
of $32$ $\MHz$, expanding our redshift coverage out to $z =
10.4$. Again, we assume that this higher redshift stretch is observed
for $t_{\rm int} = 1,000$ hours. The high redshift stretch might be
observed roughly simultaneously with the lower redshift band -- e.g.,
one day observing at low redshift followed by a day at high redshift
-- or one might observe consecutively for a year at low redshift,
followed by a year at high redshift, etc. Either way, in total our
estimates amount, optimistically, to two years of observing.

\begin{figure}
\bc
\includegraphics[width=9.2cm]{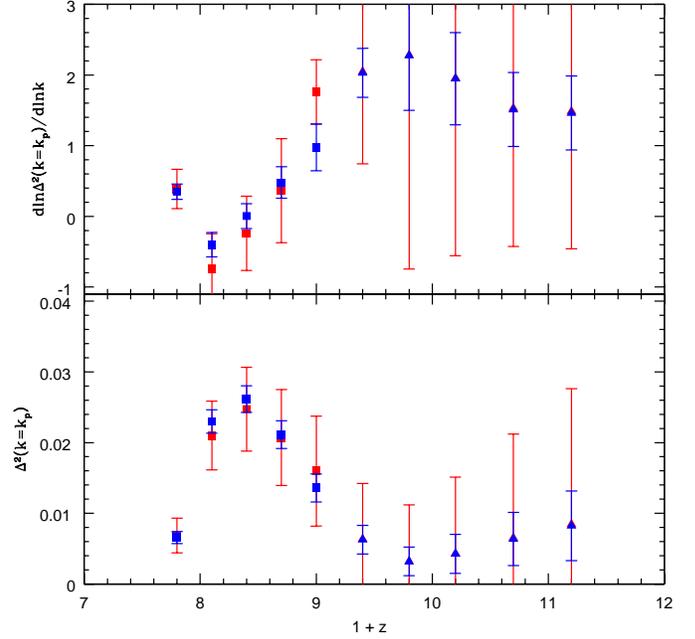}
\caption{Sensitivity of the MWA to redshift evolution in the power spectrum
amplitude and slope. 
For illustration, in each panel we consider the sensitivity of the MWA, observing for one year
(yielding $1,000$ usable hours)
over a contiguous stretch of $30$ $\MHz$, centered around a redshift of $z=7.4$ (square
points), and broken into five $6$ $\MHz$ bins. We additionally consider a separate year of observations devoted to an adjacent contiguous $30$ $\MHz$ stretch, again in $6$ $\MHz$ bins, 
but focused on higher redshifts (triangles).  
{\em Bottom}:
Redshift evolution in the power spectrum amplitude at the MWA pivot
wavenumber. The red points show the expected mean power spectrum amplitude and error bars in our fiducial
model for the MWA with an $r^{-2}$ antenna distribution, while the
blue points show forecasts for the MWA in the super-core configuration.
{\em Top}: Identical to the bottom panel, except for the slope
of the power spectrum. The slope and amplitude errors are correlated
(see text).
The MWA has the sensitivity, in the super-core configuration, to detect
the power spectrum amplitude rise and fall in this model, and the
slope flatten, with increasing redshift.
}
\label{fig:amp_slope_err}
\ec
\end{figure}

We then estimate error bars for the observing strategy described
above, using the formulas in \S \ref{sec:pspec_noise} for each of the
redshift bins considered here.  Further, we fit a polynomial function
to $\rm{ln} \Delta^2_{\rm 21}(k)$ in $\rm{ln}(k)$, truncating the fit
at the lowest order that yields an unbiased estimate of the amplitude
and slope of the power spectrum at the pivot wavenumber,
$\Delta^2_{\rm 21}(k=k_p)$:
\beqa
\rm{ln} \Delta^2_{\rm 21}(k) = \rm{ln} \Delta^2_{\rm 21}(k=k_p) + 
\sum_{j=1}^{N_p} \alpha_j \left[\rm{ln}(k/k_p)\right]^j.
\label{eq:poly_fit}
\eeqa

Since the 21 cm power spectrum is not a perfect power law on MWA
scales (Figure \ref{fig:21cm_v_z}), it is important that we leave our
fit reasonably general.  We would like to ensure that our estimates of
the amplitude and slope of the 21 cm power spectrum are unbiased: do
the best fit amplitude and slope change by more than the error bars as
we include more parameters in our fit? Further, we want to check
whether the MWA is able to detect additional parameters in the
generalized polynomial fit -- e.g., a `running' in the power law
slope. We estimate the parameters $\Delta^2(k=k_p)$, $\alpha_j$, and
their errors assuming Gaussian statistics, and using standard maximum
likelihood estimates.  We fit to the simulation data given our power
spectrum error estimates. We use all scales smaller than prohibited by
foreground contamination, and restrict ourselves to $k \leq 1 h$
Mpc$^{-1}$, since smaller scales are too noisy to yield useful
information.

We adopt $k_p = 0.4 h$ Mpc$^{-1}$ as the pivot wavenumber for our fit
(Equation \ref{eq:poly_fit}). This wavenumber is close to the middle
of the range of scales probed, and represents a convenient choice.
Note, however, that the sensitivity is a strong function of wavenumber
and the power spectrum is constrained more tightly
by the data on large scales than on small scales (Figure \ref{fig:pspec_err_v_z}).
This implies that estimates of the power spectrum amplitude
and slope are correlated for our choice of $k_p$, in contrast to 
the traditional
choice of pivot scale in which amplitude and slope errors are uncorrelated.
Provided we account for correlations in the amplitude and slope estimates,
our final constraints on the ionization 
fraction (\S \ref{sec:fill_hii}) 
are, however, independent of $k_p$. The wave number at which the slope and
amplitude errors are uncorrelated will depend on the foreground cut wavenumber,
and on redshift. 
We hence prefer to plot results in the middle of the range of scales probed
by the MWA, where the amplitude is a particularly strong function of
ionization fraction, keeping in mind the error correlations.
Moreover, note that the total $S/N$ for
21 cm power spectrum detection is {\em larger} than the significance
at which the MWA can estimate the power spectrum amplitude at this
particular scale (see \S \ref{sec:sens_results}), owing to the strong
scale dependence of thermal noise.

The results of our calculation are shown in Figure
\ref{fig:amp_slope_err}, for each of an $r^{-2}$ antenna distribution,
and the super-core configuration. In each redshift bin, as we detail
shortly, we choose the minimum number of parameters in our fit
(Equation \ref{eq:poly_fit}) that yields an unbiased estimate of the 21
cm power spectrum amplitude and slope.  Considering first the $r^{-2}$
antenna distribution, the bottom panel of the figure shows that the
MWA, in this configuration, can only weakly detect evolution in the
power spectrum amplitude at $k_p = 0.4 h$ Mpc$^{-1}$ in our fiducial
model. Still more marginal is the ability of the MWA, in the $r^{-2}$
arrangement, to detect redshift evolution in the slope of the 21 cm
power spectrum (top panel): one can almost draw a straight line
through all of the error bars in this figure. The MWA, in the
super-core configuration, however, enables significant detections of
redshift evolution in both the slope and amplitude of the 21 cm power
spectrum. 
As remarked previously, the slope and amplitude errors are
significantly correlated. For example, the correlation coefficient
between our amplitude and slope estimates is $r=0.7$
at $z=7.4$ in the super-core configuration, and $r=0.9$ at the same
redshift for the $r^{-2}$ antenna configuration.  

\begin{figure}
\bc
\includegraphics[width=9.2cm]{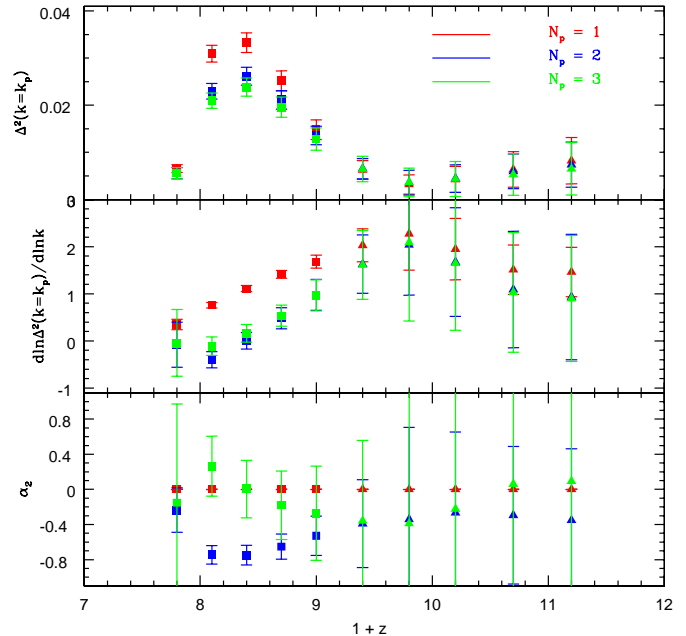}
\caption{Number of parameters constrained by MWA power spectrum measurements in 
the super-core configuration.  
We assume that $\rm{ln} \Delta^2_{\rm 21}(k)$ is a polynomial
in $\rm{ln}(k)$ and examine how many terms in the polynomial expansion the MWA can
constrain. {\em Top}: Constraints on the amplitude of the 21 cm power spectrum
at $k=k_p$. The red points show constraints assuming that $\Delta^2_{\rm 21}(k)$ is a 
power law
in $k$ over MWA-detected scales. The blue points include a quadratic fitting term 
in $\rm{ln}(k)$,
while the green points include a cubic term in $\rm{ln}(k)$. 
{\em Middle}: Similar for the slope of the power spectrum at $k=k_p$.
{\em Bottom}: Constraints on the running of the slope at $k=k_p$. At high redshift, fitting
a mere power law results in unbiased estimates of the amplitude and slope. At moderate redshift,
where the sensitivity is superior, yet there is still enough neutral hydrogen around to produce
a significant 21 cm signal ($1 + z = 8.1 - 9$ in our model), one must 
include a running in the fit to ensure unbiased
estimates of the power spectrum amplitude and slope. In these redshift bins, there appears to
be a constraint on the running, $\alpha_2$, but it is biased, as one can see by comparing the
running estimates for $N_p=2$ (blue points) and $N_p=3$ (green points). Hence, although we 
include $\alpha_2$
in the fit for these redshift bins, this only ensures unbiased constraints on the slope and 
amplitude.
}
\label{fig:params_const_v_z}
\ec
\end{figure}

Before more closely considering the insights that can be gleaned from
detecting this redshift evolution, let us briefly return to the
question of the number of parameters required.  We fit our model power
spectra with plausible MWA error-bars in the super-core configuration,
using a polynomial fit with each of $N_p = 1, 2$, and $3$. Similar
considerations apply for the $r^{-2}$ antenna distribution.  The
results of these fits are shown in Figure \ref{fig:params_const_v_z}:
the red squares show fits with $N_p=1$ included, the blue squares show
fits with $N_p=2$, and the green squares show fits with $N_p=3$. The
top panel shows constraints on power spectrum amplitude, the middle
panel shows constraints on the slope, and the bottom panel shows
constraints on the `running' of the slope, $\alpha_2$. 

We judge our estimate of a parameter `unbiased' if the preferred value
of that parameter changes by less than $1-\sigma$ when an additional
parameter is included in the fit. For example, our estimate of the
power spectrum amplitude near $1 + z \sim 8$ changes by more than one
sigma when we move from $N_p = 1$ to $N_p =2$, and we consider it
biased.  Likewise, estimates of the slope of the 21 cm power spectrum
are biased in several redshift bins for $N_p=1$. For the slope and the
amplitude, estimates converge in all redshift bins by $N_p = 2$. In
{\em each} redshift bin we include the minimum number of parameters
required to provide an unbiased estimate of the amplitude and slope,
so we include only parameters up to $N_p=1$ in some redshift bins,
while we include parameters up to $N_p=2$ in others. In some of our
most tightly constrained redshift bins there initially appears to be a
constraint on $\alpha_2$. Including an additional parameter in the
fit, $\alpha_3$, shows that our constraint on $\alpha_2$ is biased for
$N_p=2$.  Hence, while we include $\alpha_2$ as a parameter in some
redshift bins, we only use our constraints on the amplitude and slope,
since only these parameters are unbiased.

Provided the MWA packs enough antennae into its core, it does have the
sensitivity to detect redshift evolution in the slope and amplitude of
the 21 cm power spectrum, if not higher terms in the polynomial fit of
Equation (\ref{eq:poly_fit}). While our forecasts are inevitably
idealized, the constraints are tight enough that we expect a
reasonable detection of redshift evolution for our fiducial model in
the super-core configuration.

Of course, reionization may be more extended and/or occur at higher
redshift than in our fiducial model. The calculations of \S
\ref{sec:pspec_noise} give some insight into the impact of the timing
of reionization on MWA detection sensitivity.  For example, Figure
\ref{fig:variance_x_0.5_v_z} implies that we still expect a
significant detection of 21 cm fluctuations if reionization is already
$50 \%$ complete at the highest redshift probed by our hypothetical
MWA survey. In this scenario, however, the MWA would observe only the
falling half of our anticipated trend of 21 cm power spectrum
amplitude with redshift, and miss the earlier phase in which we expect
the 21 cm power spectrum amplitude to grow with decreasing
redshift. Moreover, reionization may be more extended than in our
fiducial model and lead to a more gradual power spectrum maximum when
fluctuations are measured as a function of redshift.

\section{The Filling Factor of HII Regions from Power Spectrum 
Measurements}
\label{sec:fill_hii}

Let us turn our error estimates on the 21 cm power spectrum amplitude
and slope into constraints on the filling factor of HII regions during
reionization. As an illustration, we focus on power spectrum
measurements in the specific redshift bin where the amplitude is
maximal -- i.e., our input `true' model is our fiducial model with 
$\avg{x_i} = 0.5$.  
We compare our simulated signal in this redshift bin,
including MWA statistical error estimates, with fiducial model
predictions at different ionization fractions. 
We compute $\Delta \chi^2$ between the different models, and calculate 
the likelihood
that the data are drawn from a fiducial model with a given ionization
fraction, assuming Gaussian statistics, $L = \rm{exp}(-\Delta
\chi^2/2)$.  
We assume in these calculations that our model 21 cm
power spectra are entirely fixed by $\avg{x_i}$ and ignore the weak
redshift dependence expected. Since our estimates of the amplitude
and slope are correlated, we use the full co-variance matrix to compute
$\Delta \chi^2$.

We assume our fiducial model to calculate power spectra at different
ionization fractions, ignoring model uncertainties in 21 cm power
spectra at a given ionization fraction.  This is justified -- at least
for the model of ionization fraction evolution we assume here -- since
the MWA can rule out the rare-source and mini-halo models by comparing
measurements over the full redshift range. Specifically, we translate
the power spectrum amplitude and slope measurements of Figure
\ref{fig:amp_slope_err} from measurements as a function of redshift to
measurements as a function of ionization fraction, assuming our
fiducial model to convert between redshift and ionization 
fraction.
We then compare our mock 21 cm power spectrum amplitude and slope
measurements with the rare source and mini-halo models, computing
$\Delta \chi^2$ between these models and our fiducial model. We find
that the MWA can distinguish between these models at very high
significance -- both the mini-halo and rare source model are ruled out
at $\gg 3-\sigma$ confidence even with the $r^{-2}$ antenna
configuration.\footnote{In detail, we should marginalize over the uncertain 
$\avg{x_i}(z)$ relation -- which is what we ultimately aim to constrain --
when constraining the rare source and mini-halo models.
For simplicity, we assume the fiducial $\avg{x_i}(z)$ relation in comparing
with the rare source and mini-halo models, and
find that these models are very strongly ruled out. From Figure
\ref{fig:amp_v_x}, we believe these constraints
are strong enough that they would not be alleviated by marginalizing over
$\avg{x_i}(z)$.}
This means that the MWA can break degeneracies between
the different models, and we can get approximate error estimates on
the ionization fraction from our fiducial model alone. 

\begin{figure}
\bc
\includegraphics[width=9.2cm]{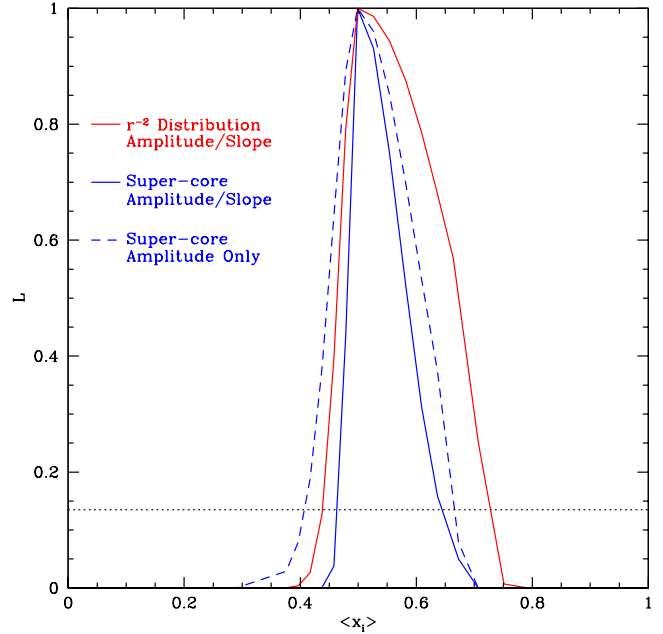}
\caption{Ability of the MWA to constrain the ionization fraction in our fiducial
model. The curves show calculations of the likelihood that an MWA measurement of
the amplitude and slope of the 21 cm power spectrum, at a redshift where the
power spectrum amplitude is maximal, is drawn from a model with a given (volume-weighted)
ionization fraction. The red solid curve assumes an $r^{-2}$ antenna
distribution and uses measurements of both the power spectrum amplitude and slope. The 
narrower
blue solid curve shows the anticipated constraint for the MWA in the super-core configuration,
using information from both the slope and the amplitude of the 21 cm power spectrum.
The blue dashed curve is similar but shows the likelihood function obtained from the
amplitude alone. The black horizontal dotted line shows the $2-\sigma$ likelihood level.
}
\label{fig:like_v_x}
\ec
\end{figure}

The results of our ionization fraction likelihood calculations are
shown in Figure \ref{fig:like_v_x}.  The MWA, observing in the
$r^{-2}$ configuration, will provide a modest constraint on
the ionization fraction at maximal amplitude: ionization fractions
between $\avg{x_i} = 0.4-0.75$ are allowed at $2-\sigma$ confidence.
The constraint is only modest because the slope and amplitude measurements
have rather large error bars in this configuration (Figure
\ref{fig:like_v_x}), and because the amplitude and slope are fairly
flat functions of ionization fraction around the peak (Figure
\ref{fig:amp_v_x}). The constraint is tighter at
low ionization fraction than one
might naively guess from eye-balling the results of Figure \ref{fig:amp_v_x},
however, since the slope and amplitude errors are correlated. 
In particular, data drawn from our low ionization fraction models will tend to have
a smaller amplitude, yet a steeper slope, than data drawn from our $\avg{x_i}=0.5$
model. Properly accounting for correlations between our slope and amplitude estimates 
then
increases the constraint on low ionization fraction models, compared to ignoring these
correlations.

On the other hand, the MWA observing in the super-core configuration
provides a bit tighter constraint. Only $\avg{x_i} = 0.45-0.65$ is
allowed at $2-\sigma$ confidence. Comparing the blue solid and dashed
curves in the figure illustrates that most of the constraint comes
from the amplitude measurement and not the slope measurement. 
Provided the middle phase of reionization occurs in the MWA observing
band, a couple of years of observations should constrain $\avg{x_i}$
around $\avg{x_i}=0.5$ to within roughly $\pm \delta \avg{x_i} \sim 0.1$ at
$2-\sigma$ confidence.

\section{Conclusions} \label{sec:conclusions}

In this paper, we considered the sensitivity of the MWA for 21 cm
power spectrum measurements and examined the resulting insights into
reionization.  Rather generically, the 21 cm power spectrum on MWA
scales increases in amplitude with decreasing redshift, as HII regions
grow and boost the level of large-scale fluctuations in the 21 cm
signal, until the time at which $\sim 50\%$ of the volume of the IGM
is ionized. Subsequently, later in the EoR, the amplitude of the 21 cm
power spectrum drops as neutral hydrogen becomes scarce. In
conjunction, the slope of the 21 cm power spectrum flattens as HII
regions grow and the 21 cm power spectrum transitions, on the scales
of interest, from simply tracing density fluctuations to tracing
fluctuations in the ionization field.

These generic features imply that, although MWA measurements will be
limited to a dynamic range of $\sim$ a decade in scale, the experiment
can significantly constrain reionization by comparing measurements in
several redshift bins. In particular, the MWA may detect the power
spectrum amplitude rise with decreasing redshift, before subsequently
turning-over and dropping in amplitude, with the slope of the power
spectrum flattening in conjunction.  This behavior is a signature of
the IGM passing through a redshift where $\sim 50\%$ of its volume is
filled with ionized bubbles.

It is unlikely that residual foreground contamination would share the
characteristic redshift evolution in power spectrum amplitude and slope 
found in our models. Measuring
this characteristic redshift evolution will hence help confirm that 
detected 21 cm fluctuations originate
from the high redshift IGM, and cement the case for neutral material in 
the high redshift IGM. 

We argued that the MWA is sensitive enough to detect this
characteristic redshift evolution in the 21 cm power spectrum
amplitude and slope, especially if it adopts a very compact configuration
for its antenna tiles, and that reionization occurs at sufficiently
moderate redshifts. A compact configuration optimizes power spectrum
sensitivity because proposed configurations, with antennae distributed
out to large $\sim 1$ km distances, populate long baselines too
sparsely to be useful.  Some antenna tiles are needed at long
baselines for point source detection and antenna calibration, but we
estimate that these requirements are not severe. 

Another possible
advantage of the super-core configuration relates to its well-behaved beam.
A potential difficulty for 21 cm observations is that point sources far
from the primary beam of an interferometer can enter through the beam's sidelobes,
which are frequency dependent. This contamination will have structure in frequency
space, and escape the usual foreground removal strategies (e.g. Oh \& Mack 2003, 
Furlanetto et al. 2006a). This contamination will be significantly reduced by adopting a very
compact antenna distribution, like the super-core configuration, which will have
a very clean beam with minimal sidelobes.

Adopting a
sufficiently compact antenna distribution, we believe that the MWA can
move past a mere detection of 21 cm fluctuations, and constrain the
volume filling factor of HII regions within two years of observing.
In particular, measurements of the 21 cm power spectrum amplitude and
slope at a redshift where $50\%$ of the volume is ionized translate
into $2-\sigma$ allowed ionization fractions of roughly $\avg{x_i} =
0.45-0.65$.  Moreover, we find that the MWA can rather easily
distinguish between our fiducial model and each of our rare source and
mini-halo models with $2,000$ hours of data, at least if the bulk of
reionization occurs in the range of redshifts probed by the MWA.

In this paper, we focused on the 21 cm power spectrum, but
investigating other statistical measures would be interesting.  One of
the primary goals of reionization studies is to use observational
measures to constrain the sizes and filling factor of HII regions
during reionization. Clearly the ionization field, and the 21 cm
signal, are highly non-Gaussian in the EoR, implying that there is, in
principle, more information than contained in the 21 cm power spectrum
alone.  We have argued that the redshift evolution of the 21 cm power
spectrum contains relatively robust information regarding the filling
factor of HII regions during reionization, but these constraints may
not be optimal and are rather indirect.  On the other hand, it is
unclear how beneficial higher order statistics will be in the low
$S/N$ regime relevant for the MWA and other first generation
surveys. We plan further investigation regarding the utility of
various non-Gaussian statistical measures, considered as a function of
instrumental sensitivity. Particularly interesting is that some
low-$k$ bins become sample-variance dominated in the super-core configuration,
implying that the MWA can actually {\em image} large scale modes in this
configuration.

Another useful endeavor would be to perform a mock MWA simulation,
incorporating thermal noise, foreground models, and the MWA
instrumental response and observing strategy.  
Our forecasts are inevitably simplified in
ignoring these details, and likely optimistic in this regard. Mock
simulations should quantify how much the sensitivity of the MWA is
degraded when incorporating such real world details. Moreover, it
would be interesting to examine how possible residuals from incomplete
foreground cleaning might bias our constraints. Since the power spectrum
variance increases strongly with wavenumber, the total $S/N$ at which the MWA
can detect the 21 cm power spectrum depends strongly on the precise
foreground cut. This is an important issue for further investigation.

Finally, in this paper we focused on a single parameter, the filling
factor of ionized regions at different redshifts, and gave simple
arguments for how one can constrain this quantity with the redshift
evolution of the 21 cm power spectrum alone. A complementary approach
for forecasting MWA parameter constraints would be to perform 
a multi-parameter Fisher matrix analysis.  

We anticipate that the MWA will swiftly move beyond a mere detection of
21 cm emission from the high redshift IGM, and obtain valuable
insights regarding the filling factor of HII regions at different
stages of reionization.

\section*{Acknowledgments}
We are grateful to Suvendra Dutta for help with the simulations
used in this analysis, and for useful discussions.
We thank Miguel Morales for comments on a draft, and for informative 
discussions, and Judd Bowman and Peng Oh for helpful conversations.
The authors are supported by the David and Lucile Packard Foundation, the 
Alfred P. Sloan
Foundation, and grants AST-0506556 and NNG05GJ40G. OZ thanks the Berkeley Center for Cosmological Physics
for support.



\end{document}